# Closing the Motivation Gap: Incentives Enhance Visual Misinformation Discernment and Verification

**Sijia Qian[1*], Cuihua (Cindy) Shen[2], Jingwen Zhang[3], Magdalena Wojcieszak[4]**

[1] Assistant Professor, School of Data Science, University of North Carolina at Charlotte, 9201 University City Blvd, Charlotte, NC, 28223, USA; Email: sqian@charlotte.edu; https://orcid.org/0000-0001-7949-1153

[2] Professor, Department of Communication, University of California, Davis, 1 Shields Avenue, Davis, CA 95616, USA; e-mail: shencuihua@gmail.com

[3] Associate Professor, Department of Communication and Department of Public Health Sciences, University of California, Davis, 1 Shields Avenue, Davis, CA 95616, USA; e-mail: jwzzhang@ucdavis.edu

[4] Professor, Department of Communication, University of California, Davis, 1 Shields Avenue, Davis, CA 95616, USA; Associate Researcher, Center for Excellence in Social Science, University of Warsaw; e-mail: mwojcieszak@ucdavis.edu

**Funding**

The authors gratefully acknowledge funding support from Facebook's Foundational Integrity & Impact Research Grant and from the European Research Council (ERC) via a Consolidator Grant (No. 101126218, *NEWSUSE: Incentivizing Citizen Exposure to Quality News Online: Framework and Tools*). Any opinions, findings, conclusions, or recommendations expressed in this material are those of the author(s) and do not necessarily reflect the views of Meta or the European Research Council.

**Competing Interests Statement**

The authors declare no competing interests.

## Abstract

Cheapfakes, or real images presented misleadingly or in unrelated contexts, are an increasingly prominent form of visual misinformation. While media literacy interventions can enhance individuals' ability to detect such content, motivational barriers often hinder the adoption of image verification. This study examines whether incorporating different mechanisms and types of incentives into a digital media literacy intervention improves visual misinformation discernment and image verification behavior, both immediately and over time. We conducted a pre-registered two-wave between-subjects online experiment (N = 1,421) on a professionally designed social media platform. The study used a 2 (Incentive Type: symbolic vs. monetary) × 2 (Incentive Mechanism: task- vs. result-based) factorial design with additional control groups. Results show that task-based incentives, particularly monetary ones, were most effective at initiating image verification behaviors (i.e., reverse image search) and boosting short-term discernment, whereas result-based incentives were more effective in sustaining discernment accuracy. These findings suggest that both the mechanism and the type of incentives play a critical role in shaping the short- and long-term effectiveness of media literacy interventions, highlighting the value of multi-phased incentive strategies for combating visual misinformation in digital environments.

## Introduction

False information, increasingly presented in multimodal media formats integrating text, video, image, and audio, poses a great threat to individuals and society at large (1–3). Among its various forms, cheapfakes – or visual content presented misleadingly or in unrelated contexts - are increasingly common sources of misinformation (4, 5). Such content is sometimes perceived as more credible than deepfakes or AI-generated synthetic content (6) and – as such - can foment misinformation endorsement, with its negative downstream effects on attitudes or behaviors (7).

Media literacy interventions are a promising way to combat misinformation, also that in visual formats (8–10). Such interventions aim to enhance an individual's ability to effectively navigate the information landscape by providing necessary skills and knowledge (11, 12). A key aspect of media literacy involves verification, which entails proactive actions to search and confirm whether a message is true, misleading, or false (13). Media literacy interventions often teach verification strategies such as lateral reading (cross-checking information by opening new tabs) and reverse image searches (tracing the origin and authenticity of images) (10, 14). While these tools are easily accessible, many people do not consistently use these verification strategies in practice. Research shows that even individuals trained to evaluate online sources often fail to apply these skills without explicit prompting (15). Thus, although media literacy interventions can successfully build verification skills, a persistent gap remains between skill acquisition and behavior adoption, highlighting motivation as a crucial but often overlooked barrier to effective misinformation mitigation.

One way to boost motivation is to use incentives (16). By providing tangible financial rewards or recognition, individuals can be motivated and habitualized to utilize tools designed to validate information accuracy and act upon the learned knowledge and skills. Indeed, incentive-based programs are successful in similar challenging behavior domains, such as vaccination campaigns and pro-environment initiatives (17, 18). Integrating incentives into digital platforms and tools may not only enhance their effectiveness but also cultivate a persistent habit of critical information assessment. This can ultimately contribute to a new norm of information engagement patterns, reducing the spread of misinformation (19).

Incentives can be monetary or symbolic. Prior research finds mixed evidence on the effectiveness of monetary incentives in countering misinformation. Whereas some work shows that providing a cash bonus for accurate judgments improves participants' ability to discern true vs. false news headlines (20), reduces partisan bias in those judgments (21), and promotes the sharing of true information (22), other studies find that offering a bonus for being accurate does not significantly boost discernment between true and false statements or reduce the repetition-induced truth effect (23). In short, monetary incentives might have varying effectiveness and – furthermore – may lose effectiveness over time (24).

In contrast, symbolic incentives refer to non-monetary rewards, such as digital badges. Although they carry no direct financial value, they tap into psychological drivers like a sense of accomplishment, social status, or altruistic fulfillment (25). Compared to monetary incentives, symbolic incentives may foster more sustained engagement with the task by promoting intrinsic motivation and deeper emotional investment of an individual (26). This enduring connection can persist even after the incentives are removed (27). Symbolic incentives are also low-cost and easily scalable, making them suitable for wide implementation. Despite their potential promises, empirical evidence remains limited on whether such cost-free symbolic rewards alone are sufficient in the misinformation context in the long term.

Rewards differ not only in type but also in how they are earned (i.e., incentive mechanism), a core factor that is often overlooked. Task-based incentives are rewards given for merely completing a task regardless of the result (i.e., reverse searching image posts in our study), while result-based incentives provide immediate feedback and rewards based on the performance (i.e., accurate evaluations of the credibility of image posts). These two incentive mechanisms shape motivation in distinct ways and lead to different behavioral outcomes. Result-based rewards tend to elicit greater effort and often lead to higher output quality and longer engagement (28). In contrast, task-based incentives are more effective for encouraging initial participation and reducing barriers to entry, making them useful for promoting widespread use (26). Over time, they may also foster sustained involvement by allowing individuals to build competences without the pressure of trying to be accurate based on immediate evaluation, supporting long-term learning and skill development (29). In the context of misinformation verification, comparing these incentive strategies can help identify whether motivating accurate result or motivating task completion yields more enduring and scalable behavior change, and outcome of key practical and also theoretical importance.

Despite the potential large-scale implications of diverse incentive mechanism and incentive type in anti-misinformation and media literacy interventions more broadly, we lack at-scale evidence on their over time effects. This study aims to fill this critical gap to investigate which combinations of incentive mechanism and type most strongly enhance the effectiveness of a digital media literacy intervention over both the short term (i.e., immediately following the incentives) and long term (i.e., when incentives are no longer present). We focus on two key outcomes: individual ability to discern visual misinformation and the intention to adopt reverse image search verification. By measuring both perceptual accuracy and behavioral intent, the study offers a comprehensive evaluation of the effects of incentive strategies.

We report the results from a two-wave online experiment (N= 1,419) using a 2 (incentive mechanism: result- vs. task-based) by 2 (incentive type: monetary vs. symbolic) factorial design, plus a no-incentive intervention group and a separate control group that received no intervention (Figure 1). Participants were recruited from Prolific using quota sampling to approximate U.S. census distributions. In Wave 1, participants were directed to a professionally designed, interactive social media platform that simulated a Facebook-like environment (Figure 2). After completing baseline measures, participants in the intervention conditions viewed a digital media literacy infographic teaching reverse image search (Figure 3), whereas the no-intervention control group viewed an unrelated infographic. Participants then evaluated four news posts. These posts consisted of authentic, unedited images paired with captions adapted from fact-checking articles covering diverse topics on Snopes, either accurately describing the depicted event or presenting an unrelated context that rendered the image misleading; half of the posts were accurate, and half contained out-of-context visuals. After each post, participants rated its accuracy. Incentives were embedded within this task. In the monetary condition, participants received payment bonus, immediately shown via a pop-up window and redeemable in the end. In the symbolic condition, participants earned a verification badge that was instantly displayed on their profile. Incentives were either task-based (rewarded for conducting a reverse image search) or result-based (rewarded for correctly identifying post accuracy). Following the task, participants reported their intention to use reverse image search in the future. One week later, participants were recontacted for Wave 2 (final N = 1,147; retention rate = 80.8%). They evaluated the same four posts from Wave 1 along with two new posts (one accurate and one misattributed), again rating accuracy. No incentives were provided in Wave 2, allowing us to assess the persistence of intervention and incentive effects over time.

Overall, we find that incentives do meaningfully reduce the motivation gap in visual misinformation contexts, yet their effectiveness depends critically on both incentive mechanism and type. Task-based monetary incentives are most effective at initiating verification behaviors and boosting short-term discernment, whereas result-based incentives, both symbolic and monetary, are more effective at sustaining discernment once incentives are removed. These findings demonstrate the potential of incorporating incentives in media literacy interventions to move beyond skill acquisition and actively promote verification behavior and evaluative accuracy over time. Our results also offer a more optimistic perspective on user-centered misinformation mitigation by showing that even modest, scalable rewards—when strategically designed—can foster both verification engagement and lasting cognitive gains in real-world–like social media environments.

## Results

Full details on the experimental design and study procedures are available in the Materials and Methods section and Supporting Information (SI). In SI Section 1, we offer details on participant recruitment, the demographic composition of our sample (see SI Table S1), well as evidence that attrition did not differ by experimental condition.

### Optimal Combination: Getting Money for Completing the Task

We first examine which combination of incentive strategies, including both incentive mechanism and type, is most effective in enhancing participants' ability to discern visual misinformation (discernment) and their intention to use reverse image search to verify visual content (verification intention). The question wording for these items is included in the Materials and Methods and SI Section 3. SI Table S4 presents descriptive means and standard errors for discernment and verification intentions across experimental conditions at Wave 1 and Wave 2. Overall, task-based monetary incentive works best when compared to both the no-incentive group and other incentive groups. Figure 4 displays mean discernment scores across treatment groups. The task-based monetary incentive condition exhibited the highest discernment in Wave 1 (M = 0.25), compared to the task-based symbolic (M = 0.07), result-based symbolic (M = –0.21), result-based monetary (M = 0.02), and no-incentive control groups (M = –0.04). This represents a 0.29-point increase relative to the control group on the –4 to 4 discernment scale.

Participants in the task-based monetary group also indicated the highest level of intention to use reverse-search verification in the future (M = 4.34; on a five-point scale) compared to the other incentive groups (task-based symbolic group: M = 4.09; result-based symbolic group: M = 3.93, result-based monetary group: M = 4.11) and the no incentive control (M = 3.94) in Wave 1. Compared to the control, we observe a 10% increase in future intention to use reverse-search in the task-based monetary group.

In Wave 2, conducted one week after the treatment, the task-based monetary group maintained a clear advantage over the other incentive groups and the control group in terms of participants' intention to use reverse image search tools in the future. Participants in this group reported the highest intention (M = 4.12 on a five-point scale), outperforming the task-based symbolic group (M = 3.89), result-based symbolic group (M = 3.85), result-based monetary group (M = 3.83), and the no incentive control group (M = 3.69).

However, the advantage of the task-based monetary group in discernment was less apparent in Wave 2. Instead, participants in result-based groups, those receiving either symbolic or monetary incentives, demonstrated the highest levels of discernment (M = 0.94 and 0.93 respectively) compared to the task-based groups. We discuss this comparison in the following sections.

**Differential Effects of Incentives on Discernment and Verification Intention**

We further investigated how different incentive mechanism (task-based vs. result-based) and incentive types (symbolic vs. monetary) shaped participants' discernment and verification intention across two waves using regression analyses, controlling for demographics, political leaning, literacy, and media trust. SI Table S3 presents the means, standard deviations, and zero-order correlations (with 95% confidence intervals) among the main outcome variables across both waves, as well as these key covariates. The full regression tables are presented in SI Table S5 and S6.

Our findings reveal a shift in the effectiveness of different incentive strategies over time (see Figure 5). In Wave 1, the only incentive strategy that significantly improved discernment was the task-based monetary incentive ($b = 0.298$, $p = .035$), suggesting that offering a direct financial reward for completing the verification task modestly enhanced participants' ability to distinguish between accurate and misleading content. This suggests that tangible, straightforward incentives can prompt participants to engage more carefully with visual content in the short term.

However, this short-term advantage did not persist over time. By Wave 2, a different pattern emerged. Rather than task-based incentive, result-based incentive, that is tied to participants' discernment performance, was more effective. Both the result-based symbolic ($b = 0.265$, $p = .031$) and result-based monetary incentives ($b = 0.244$, $p = .047$) predicted higher discernment. These findings suggest that by Wave 2, when incentives were no longer available, discernment was actually highest among those individuals who were rewarded in Wave 1 for how well they performed, not for simply completing the task. Importantly, the size of the coefficients indicates a moderate effect: participants offered result-based incentives (either a symbolic badge or a financial reward), scored approximately a quarter of a point higher on the discernment scale compared to those in the control group. Given discernment's standard deviation (1.15 in Wave 2), this translates into a small-to-moderate practical impact. This shift highlights a critical distinction between incentive mechanisms: task-based incentives may be better suited for initial engagement, while result-based incentives are more effective for building lasting cognitive skills like discernment.

In contrast, the effects of incentives on verification intention, that is participants' willingness to engage in future verification behaviors, were both stronger and more consistent across time (see Figure 5). In both waves, task-based monetary incentive consistently and significantly increased participants' intention to verify visual content (Wave 1: $b = 0.417$, $p < .001$; Wave 2: $b = 0.434$, $p < .001$). Participants in the task-based monetary incentive condition reported verification intentions that were 8.3% higher in Wave 1 and 8.7% higher in Wave 2 compared to those in the control group. These effects represent moderate and practically meaningful increases, underscoring the motivational value of even small monetary incentives. Interestingly, result-based incentives, whether symbolic or monetary, did not significantly predict verification intention in either wave. This pattern suggests that while people may be motivated to verify information when they receive an upfront incentive for participation, linking rewards to their verification accuracy may be less effective, potentially due to uncertainty about their own performance or the delayed nature of task-based rewards.

Taken together, these results indicate that task-based monetary incentives are especially effective at increasing participants' intentions to verify visual content, while result-based incentives are more useful in enhancing their actual discernment of misinformation, particularly over time.

## Discussion

Motivating people to verify online content is a perennial challenge, especially given the spread of increasingly sophisticated visual misinformation. This challenge is amplified for images as they provide fewer searchable cues and elicit faster, more heuristic processing than text (30). Addressing this motivation gap, this study examined how different incentive strategies based on incentive type (symbolic vs. monetary) and incentive mechanism (task-based vs. result-based), affect individuals' intention to engage in verification activities (i.e., reverse image search) and their ability to discern factual content from cheapfakes. Our findings provide several key insights into the efficacy and temporal dynamics of these incentive strategies and highlight broader implications for media literacy interventions and platform design.

### Money Buys Short-term Gains While Symbolic Badges Result in Lasting Impact

In Wave 1, providing monetary rewards based on task completion emerged as the most effective strategy. Participants who received monetary incentives for simply performing verification behaviors (i.e., using reverse image search tools) showed modest but statistically significant improvements in discernment, along with the strongest intentions to continue employing verification strategies in the future. This underscores the value of simple, tangible incentives to jump-start critical evaluation behaviors. This is consistent with previous literature on behavioral nudges and extrinsic motivation, which highlights that small, immediate rewards can effectively initiate new behaviors (26, 28). Yet, the sharp initial boost from monetary, task-based incentives did not persist over time. One week later (Wave 2), this group maintained a stronger intention to verify content but no longer showed discernment advantage. We speculate that the effect faded because the behavior was motivated by the reward rather than by a lasting change in how participants evaluated images. Such findings suggest that although easily attainable, task-based monetary rewards can serve as a powerful catalyst for initial engagement, they have limited power in promoting sustained accuracy in the long term.

By contrast, symbolic incentives, such as displaying a verification token next to a participant's profile, did not show as pronounced immediate effect but held promise for sustained effects on discernment. Symbolic rewards may tap into intrinsic motivators related to social recognition and personal identity, and while they did not show pronounced effects immediately, they may support more sustained engagement over time, particularly when tied to result-based mechanism (31) and especially that – arguably – such incentives are more easily implementable at scale. Although our time frame was relatively short, only one week, the diminished differences over time between monetary and symbolic groups signal that future initiatives might consider blending these approaches or phasing from monetary to symbolic incentives to maintain engagement and accuracy over longer periods.

### Task Incentives Boost Immediate Adoption While Result Incentives Sustain Accuracy

Our results also underscore the importance of distinguishing between task-based and result-based mechanisms in the context of verifying visual misinformation and in promoting digital literacy more broadly. Task-based incentives were more effective in immediately boosting verification intentions, while result-based incentives yielded longer-term accuracy benefits. These findings highlight that immediate engagement may be best achieved by simply rewarding the act of verifying potentially misleading content, while sustained improvements in accuracy are better supported by incentives tied to performance quality, namely whether individuals were correct in evaluating the verification results. This temporal complementarity suggests an evolving strategy: platforms or interventions could begin with task-based incentives to rapidly scale engagement, then gradually shift to result-based rewards that hone long-term evaluative rigor.

**Design Implications for Multi-Phased Interventions to Foster Verification**

These findings can inform the design of interventions aimed at enhancing media literacy and promoting behaviors aimed at content verification toward minimizing misinformation and increasing discernment in online information environments. Fact-checking organizations, news literacy programs, and social media platforms can leverage our results to craft multi-phased interventions. In the early stages, straightforward and tangible incentives, small monetary bonuses, might encourage users who otherwise would not bother verifying content to try out recommended tools or verification processes. Once users become familiar with the verification process, platforms could transition to result-based incentives or symbolic recognition, such as offering small badges, public acknowledgment on users' profiles, or periodic recognition in community leaderboards or featured verifier spotlights. That helps reinforcing the value of accuracy and deepening users' long-term engagement in critical evaluation. Compared to monetary rewards, symbolic incentives tied to user actions are more scalable and realistic for platforms to implement, especially given limited financial incentives and heightened political sensitivities around content moderation. Tailoring intervention to users' motivation and verification experience could also be a future direction, enabling more personalized approaches that adapt to diverse user needs (32).

The insights from our study can inform the design of lightweight features that motivate and support user verification, particularly within community-driven systems. Recent changes across major social media platforms, such as Meta and Twitter/X, mark a shift from centralized fact-checking toward community-driven models like Community Notes (33). This pivot reflects a broader move toward emphasizing user participation in misinformation verification and credibility assessments. However, the effectiveness of this model remains uncertain, with some evidence suggesting that Community Notes do not significantly reduce engagement with misinformation (34). As a result of this shift, users now play a far greater role in verification—both as individual evaluators and as contributors to various participatory systems. Yet, like most crowdsourced platforms, Community Notes face persistent challenges of uneven participation and under-contribution: only a small subset of users actively contributes, and most notes never reach publication (35). These participation gaps underscore the need for further research and investment in strategies that effectively motivate users to engage in (mis)information verification. Based on our results, platforms might initially offer users small financial rewards to boost short-term participation in fact-checking activities. Over time, as users build their verification skills, platforms could emphasize recognition tied to quality, such as providing more nuanced badges and rankings that reflect sustained correctness. Such approaches could foster a culture of critical engagement, encouraging users not just to participate, but to care about the correctness of their evaluations.

We acknowledge that our study is not without limitations. First, while one of our key outcome variables is the intention to use reverse image search, it is a measure of behavioral intention rather than of actual behavior. Prior research shows that behavioral intention is a strong predictor of future behavior (36), and so we suspect that individuals indicating their intention to use reverse search will indeed use it when faced with potential cheapfakes. Nonetheless, future work should incorporate behavioral tracking to assess whether intentions translate into real-world verification actions in naturalistic settings. Second, our sample was recruited from Prolific, a platform known to attract participants who are generally more digitally literate than the general population, which may limit the generalizability of our findings. Future studies should recruit more diverse and representative samples to examine whether the observed effects hold across different demographic and digital literacy profiles. Third, our intervention was designed for desktop or laptop use. Although reverse image search tools are technically accessible on mobile devices, they are often less user-friendly and more cumbersome to use in mobile environments. Given the growing prevalence of mobile devices as the primary means of accessing social media platforms and consuming information therein, we encourage scholars to explore how to optimize verification tools and interventions for mobile platforms to enhance usability and real-world applicability.

Despite these limitations, our findings provide compelling evidence that incentive type and incentive mechanism can meaningfully enhance the effectiveness of digital media literacy interventions. A persistent challenge for digital literacy initiatives is scalability, as individuals often fail to consistently apply newly acquired verification skills in everyday online environments (15). Against this backdrop, evidence that these incentive-based strategies not only improve discernment—the ability to distinguish between accurate information and misleading or out-of-context visual content—but also increase individuals' willingness and intention to engage in verification behaviors over time during actual platform use is both promising and consequential. Given that ongoing challenges to democratic societies are increasingly linked to the spread of online misinformation (3), identifying scalable incentive designs that promote information integrity on social media platforms is both timely and urgently needed.

## Materials and Methods

The study protocol received ethical approval from the Institutional Review Board at the [redacted for anonymous review]. The study is preregistered on OSF (view-only link: https://osf.io/t8eau/overview?view_only=fd5f81dcb4524cf4a90c25332909e6e9)

### Study Design

This study employed a two-wave, between-subjects online experiment using a 2 (Incentive Type: symbolic vs. monetary) × 2 (Incentive Mechanism: task-based vs. result-based) factorial design, plus a no-incentive intervention group and a separate control group that received no intervention. To enhance ecological validity and test how incentives affect engagement with digital verification tools in realistic settings, we developed a professionally designed, interactive in-house social media platform modeled after Facebook. On this platform, participants could browse visual posts and right-click on images to initiate reverse image searches, mirroring how such verification would occur in real online environments (see Figure 2). Participants received real-time feedback tailored to their assigned condition. For instance, participants in the monetary groups saw their earned bonus accumulate on-screen (redeemable for cash in the end). This immediate, tailored feedback was designed to reinforce the salience of the incentive strategies and strengthen the experimental manipulation.

**Sample and Procedure**

We recruited participants from Prolific, using quota sampling based on age, gender, ethnicity, and political orientation based on simplified US census data. We focus our analyses on the participants who passed attention check questions. In SI section I we report all exclusions. The final analytic sample included 1,419 participants in Wave 1. Their mean age was 42 years (SD = 14.1), 66.6% were white, and 45% identified as women. Detailed demographic sample characteristics are presented in SI Section 1. All participants provided informed consent prior to participation, in accordance with the approved study protocol. Participants were required to complete the survey using a desktop or laptop computer, as conducting a reverse image search is notably more challenging on a mobile device. Each participant was paid $2 for their time, and some participants received additional compensation as a result of their experimental assignment.

Participants were redirected to an interactive social media site modeled after Facebook. They first answered questions regarding media use, trust in media, digital media literacy, and issue relevance. Then, they were randomly assigned to one of the six experimental groups (see Figure 1), including five intervention groups and one no intervention control group: 1) task-based symbolic incentive, 2) task-based monetary Incentive, 3) result-based symbolic incentive, 4) result-based monetary incentive, 5) no incentive control, and 6) no intervention and no incentive control group. Those in the first five intervention groups were shown a digital media literacy education infographic, while those in the last control group were shown an infographic on a travel-related topic. Next, each group received a general instruction and instructions specific for receiving incentives (see SI Section 4 for details). Participants then read four news posts, each with an image and a short description (see SI Section 2 for a complete list of posts and selection process). Half of the news posts are accurate, and the other half contain out-of-context visuals. After each news post, they were asked to rate the accuracy of each post. Lastly, they answered a measure of intention to use reverse image search in the future, and other demographic questions.

In Wave 2 (one week later), the final analytical sample includes 1,147 valid responses (retention rate 80.8%). Each participant was paid $5 for their time. Participants viewed the same four posts in Wave 1 and two additional news posts (one accurately attributed and the other misattributed), and again evaluated the accuracy of each post and indicated their intention to use reverse image search in the future. In the end, participants were debriefed.

**Incentive Implementation**

We tested two types of incentives (symbolic and monetary) and two mechanisms of incentives (task-based and result-based). Monetary incentives referred to additional cash bonuses that participants could earn based on their performance or task completion during the study. Each participant viewed and evaluated four posts, and for each post that met the predefined incentive mechanism (e.g., correctly verifying an image or demonstrating accurate discernment), they received a $1 bonus, up to a maximum of $4. The total bonus amount was added to their base compensation of $2 and distributed at the end of the study. Symbolic incentives took the form of a verification badge with a green checkmark, designed to function as a status token, similar to those used on social media platforms. In the task-based conditions, participants earned incentives simply for engaging in the verification behavior, specifically, by right-clicking on image posts to

initiate reverse image searches, regardless of whether their accuracy judgments were correct. In contrast, the result-based conditions rewarded participants based on the number of posts they accurately evaluated in terms of credibility.

**Media Literacy Intervention**

The media literacy intervention, adapted from a prior study (10), targeted both knowledge and skills related to online content verification. Participants assigned to an intervention group viewed an infographic (see Figure 3) that introduced the concept of reverse image search, explained why visual misinformation is common online, and provided step-by-step guidance on how to conduct image verification. This approach aligns with established frameworks of digital media literacy, which emphasize both cognitive understanding and technical ability. By teaching manual verification techniques grounded in media literacy best practices (37), the intervention aimed to equip participants with practical, actionable skills for identifying misleading image–caption pairings.

**Key Measures**

Our primary dependent variable was discernment, operationalized as the participant-level mean difference in perceived accuracy ratings between posts with out-of-context captions and those with accurate captions, following (38). The perceived accuracy of image news posts was measured by asking participants to rate the credibility of each post using four items on a 5-point Likert scale (1 = Strongly disagree, 5 = Strongly agree). Additionally, participants reported their future verification intention by responding to the question, "How likely are you to use reverse image search tools when browsing news posts on social media in the future?" (1 = Extremely unlikely, 5 = Extremely likely). The covariates included political ideology, digital media literacy, visual literacy, media trust, social media use, and demographics, with detailed measures reported in the SI Section 3.

**Analytical Strategy**

To evaluate the effectiveness of different incentive strategies on participants' discernment and future verification intentions, we conducted both ANOVA and regression analyses. We first employed a series of one-way ANOVA tests to compare each of the four incentive strategy groups, against a no-incentive control group. These analyses allowed us to identify which specific incentive combinations were most effective in improving key outcomes.

Next, we conducted a series of ordinary least squares (OLS) regression analyses. We examined effects across two time points: immediately after the intervention (Wave 1) and one week later (Wave 2). For each wave and outcome, we estimated separate OLS models that included incentive groups as the key predictor, along with relevant covariates (e.g., age, gender, education, media use, political leaning, media literacy, visual literacy, and trust in media).

**Data Availability Statement**

Anonymized survey data and the analysis scripts used in this study are publicly available in the Open Science Framework (OSF) repository at: https://osf.io/be9tf/files/osfstorage.

**Figures and Tables**

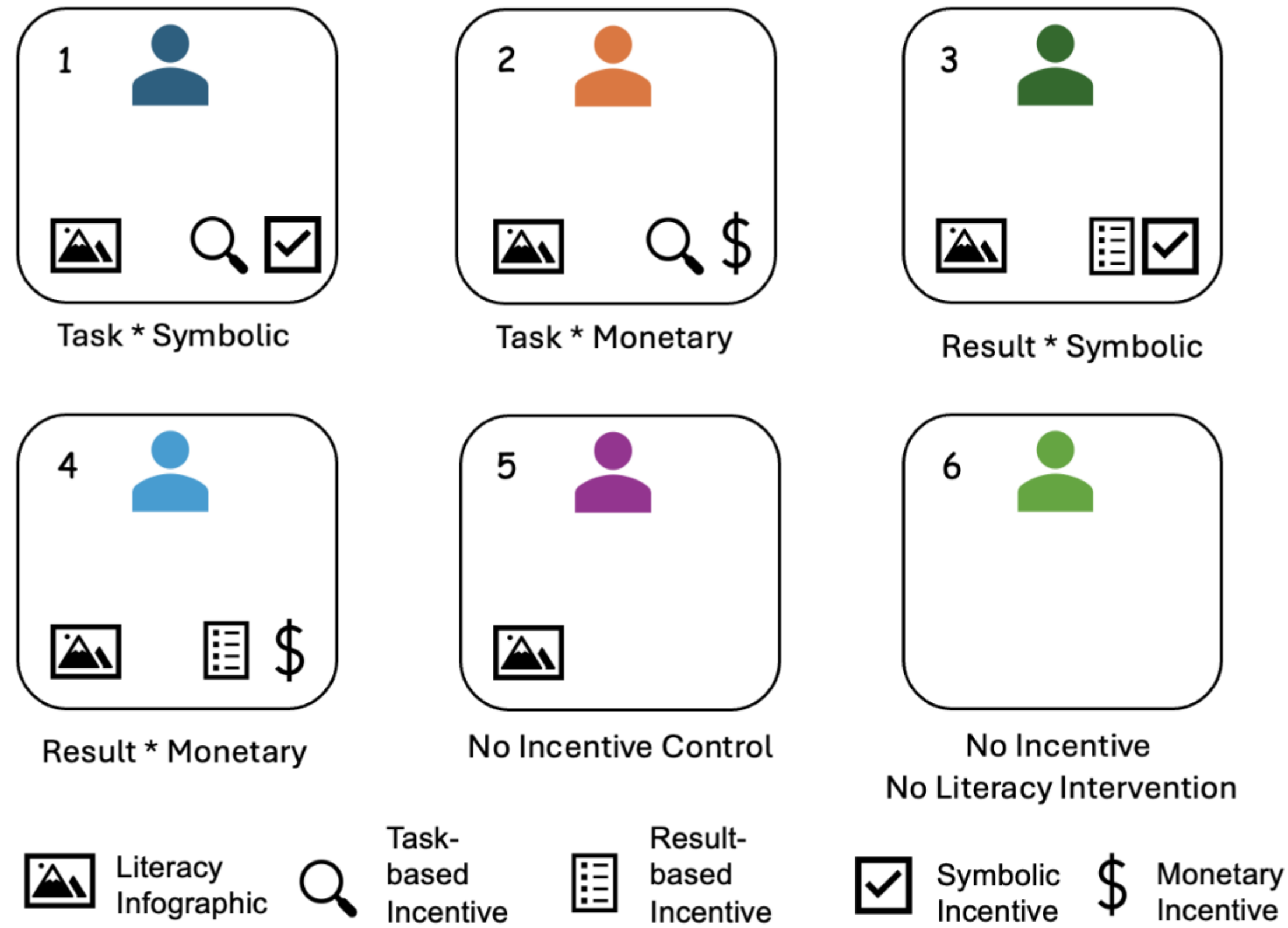

**Figure 1.** Overall design of the experiment.

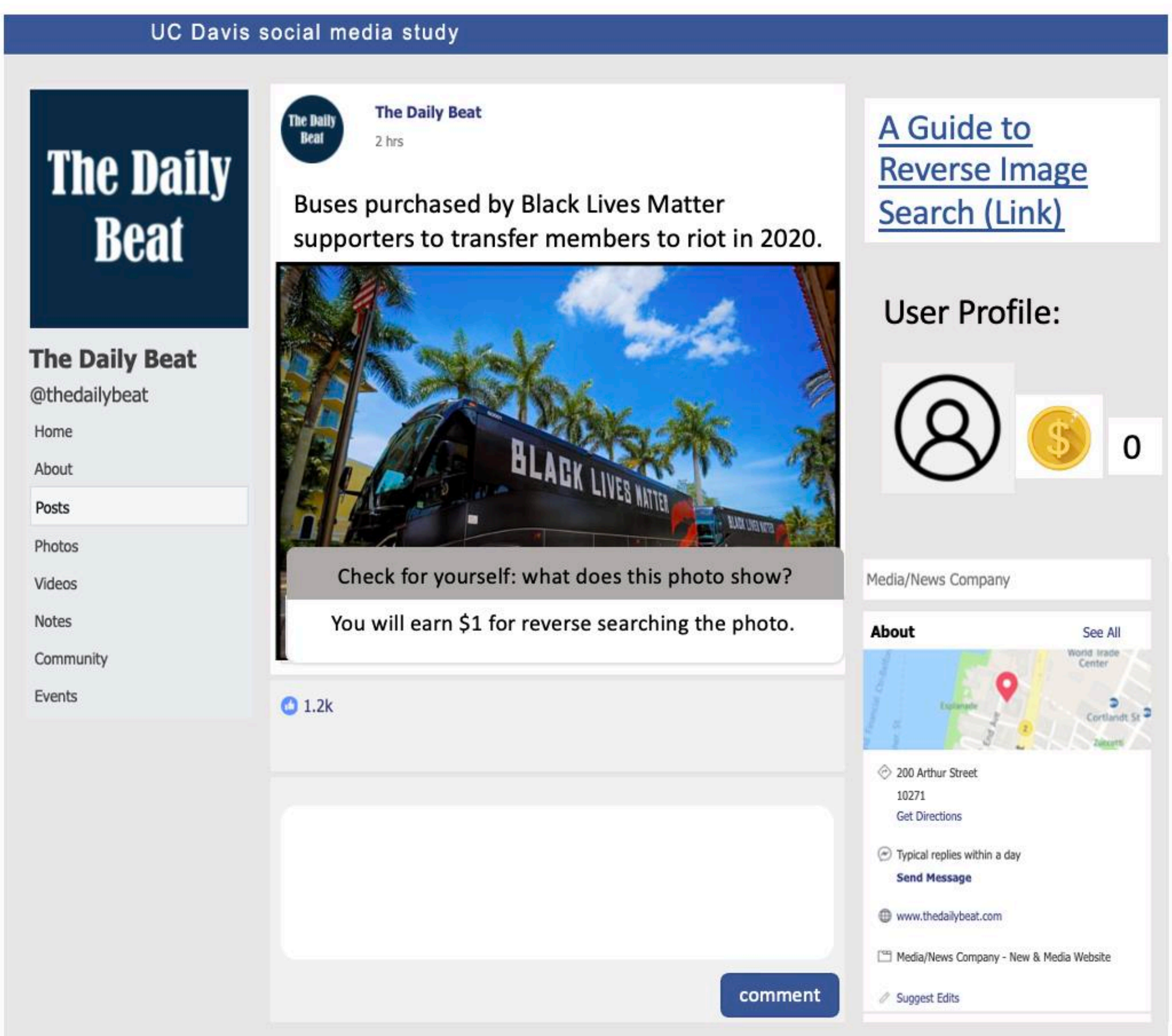

**Figure 2.** Experimental platform shown to the task-based monetary incentive group

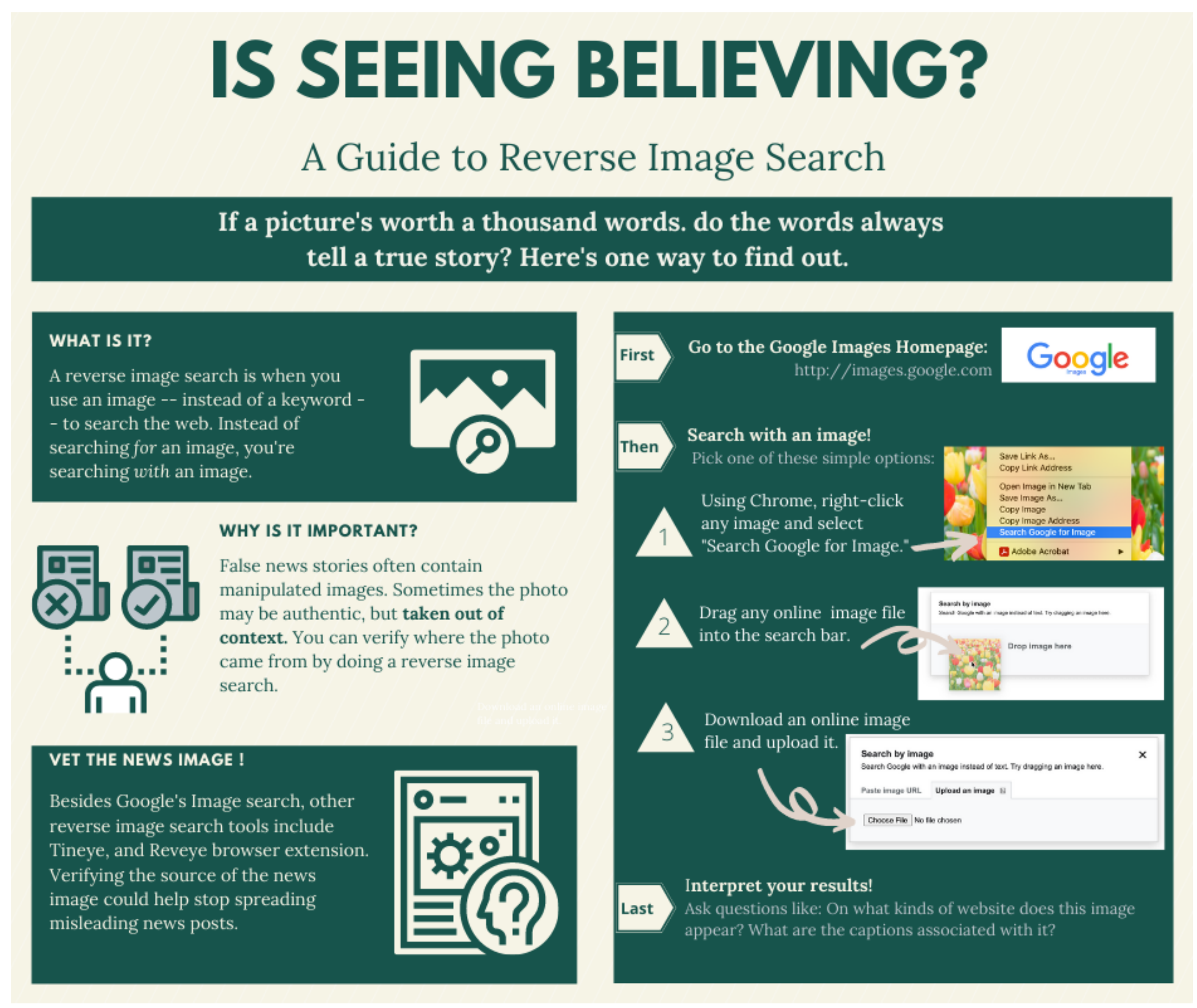

**Figure 3.** Digital media literacy infographic teaching reverse image search.

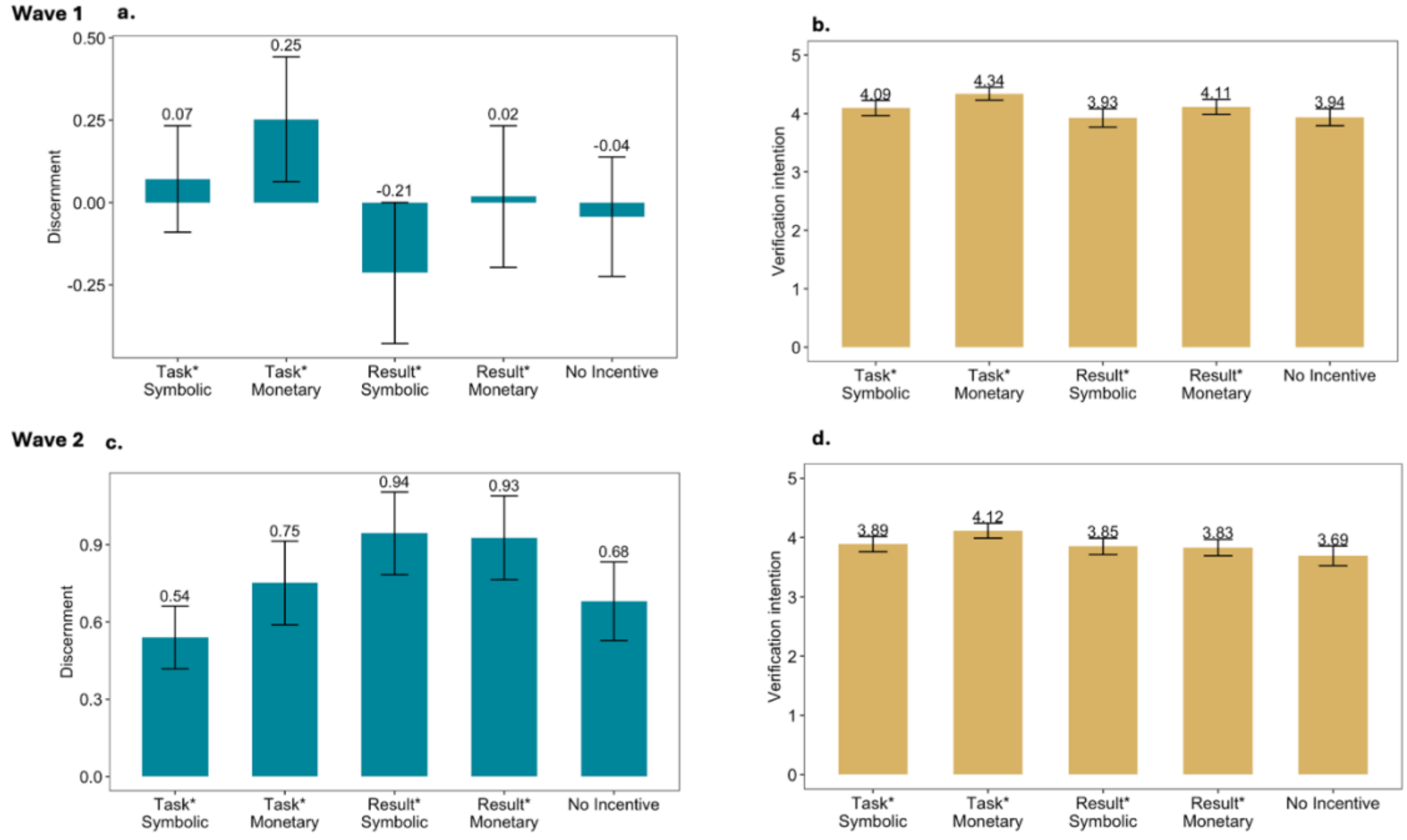

**Figure 4.** Mean discernment and verification intention by incentive strategies.

Note. This figure presents group means for discernment and verification intention across five experimental conditions: task-based symbolic, task-based monetary, result-based symbolic, result-based monetary, and no incentive. Measures were taken immediately after the intervention (Wave 1) and one week later (Wave 2). **(a)** Mean discernment scores immediately after the intervention. **(b)** Mean verification intention immediately after the intervention. (**c)** Mean discernment scores one week after the intervention. **(d)** Mean verification intention one week after the intervention. Error bars represent 95% confidence intervals around the mean.

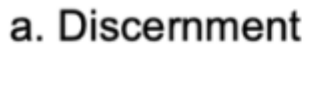

b. Verification Intention

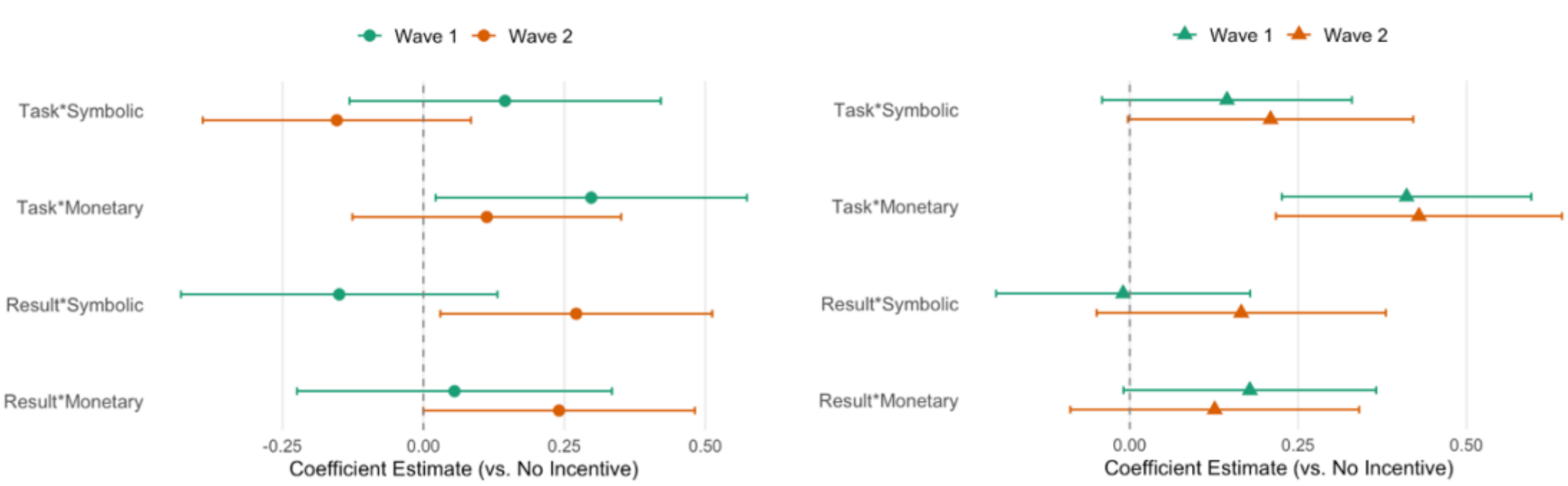

**Figure 5.** Effects of incentive strategies on discernment.

Note. This figure displays the estimated coefficients and 95% confidence intervals for each incentive condition (vs. the no-incentive control group) on discernment (left) and verification intention (right) across two waves. Horizontal lines represent coefficient estimates from Wave 1 (upper point) and Wave 2 (lower point) for each condition. Models control for age, gender, income, education, media use, political leaning, media literacy, visual literacy, and trust in mass media and social media. A dashed vertical line at zero indicates the reference point (no effect compared to control).

**Supporting Information for**

# Closing the Motivation Gap: Incentives Enhance Visual Misinformation Discernment and Verification

**This file includes:**

Supporting text
Tables S1 to S11

## Table of Contents

## Part I: Methods Supplement

### Section 1: Sample Demographics

In Wave 1, we recruited 1,490 participants. To ensure data quality, we excluded 69 participants who failed either of the two attention check questions and two participants whose reported age did not meet our inclusion criterion (18 years or older), resulting in a final Wave 1 sample of 1,419 participants. In Wave 2, 1,203 participants returned to complete the follow-up survey. After removing two participants who failed the attention check and matching responses with valid Wave 1 data, the final Wave 2 sample consisted of 1,147 participants. Table below presents the demographic characteristics of the sample.

Attrition did not differ by experimental condition, $\chi^2(5) = 1.09$, $p = .96$ or by education. Although attrition varied slightly by gender, $\chi^2(2) = 7.22$, $p = .03$, the magnitude of these differences was modest (5–9 percentage points) and unrelated to condition assignment. Attrition also varied by age, with younger participants more likely to drop out between waves (e.g., 34.2% among those aged 18–24 vs. 11.4% among those aged 65+), a pattern commonly observed in panel surveys.

**Table S1**. Demographics of Sample

| | Wave 1 (%) | Wave 2 (%) |
|---|---|---|
| Gender: Male | 52.2 | 54.4 |
| Gender: Female | 44.7 | 43.3 |
| Gender: Others | 2.1 | 2.3 |
| Age: 18-24 | 8.0 | 6.5 |
| Age: 25-34 | 28.7 | 27.4 |
| Age: 35-44 | 23.3 | 24.1 |
| Age: 45-54 | 17.3 | 17.8 |
| Age: 55-65 | 15.2 | 16.1 |
| Age: 65+ | 7.4 | 8.1 |
| Education: None, or grade 1-8 | 0.2 | 0.2 |
| Education: High school incomplete | 1.1 | 1.1 |
| Education: High school graduate | 12.2 | 12.2 |
| Education: Technical, trade or vocational school AFTER high school | 4.8 | 4.9 |
| Education: Some college completed | 25.7 | 25.5 |
| Education: College graduate | 40.7 | 39.6 |
| Education: Post-graduate training | 15.4 | 16.5 |

**Section 2: Stimuli Image News Posts**

Prior to conducting the main study, we performed two rounds of pretests at a western public university using the school's online research participation system. Undergraduate students participated in the pretests and received course credit for their involvement (n = 248 and n = 123 for the first and second rounds, respectively). The purpose of the pretests was to ensure that perceived credibility did not vary significantly among the different news contexts in our study or between accurately attributed and misattributed visual posts. This was intended to minimize the influence of image selection on the study's results.

We created a total of 28 visual posts based on 14 images (14 images × 2 captions: accurately captioned vs. misattributed). Each visual post consisted of one real, unedited image accompanied by a text caption sourced from fact-checking articles on Snopes.com. The captions either accurately described the events depicted in the image or presented a completely unrelated scenario, rendering the image out of context. All posts were designed using a generic social media post format without platform logos, considering that individuals likely consume news using different social media sites. To avoid biased processing, the source of the post was presented using a fictitious news outlet, *The Daily Beat*, which was created for this study to resemble a real but neutral news site. Eight images were used in the first round, and six were used in the second round. During the pretests, each participant rated the credibility of the posts. From all the posts, we selected six images that had similar credibility ratings across news contexts and between accurately attributed and misattributed conditions. Four images served as stimuli in Wave 1, and in Wave 2, participants were shown these four again along with two additional images.

**Table S2.** Stimuli Used in Wave 1 and Wave 2.

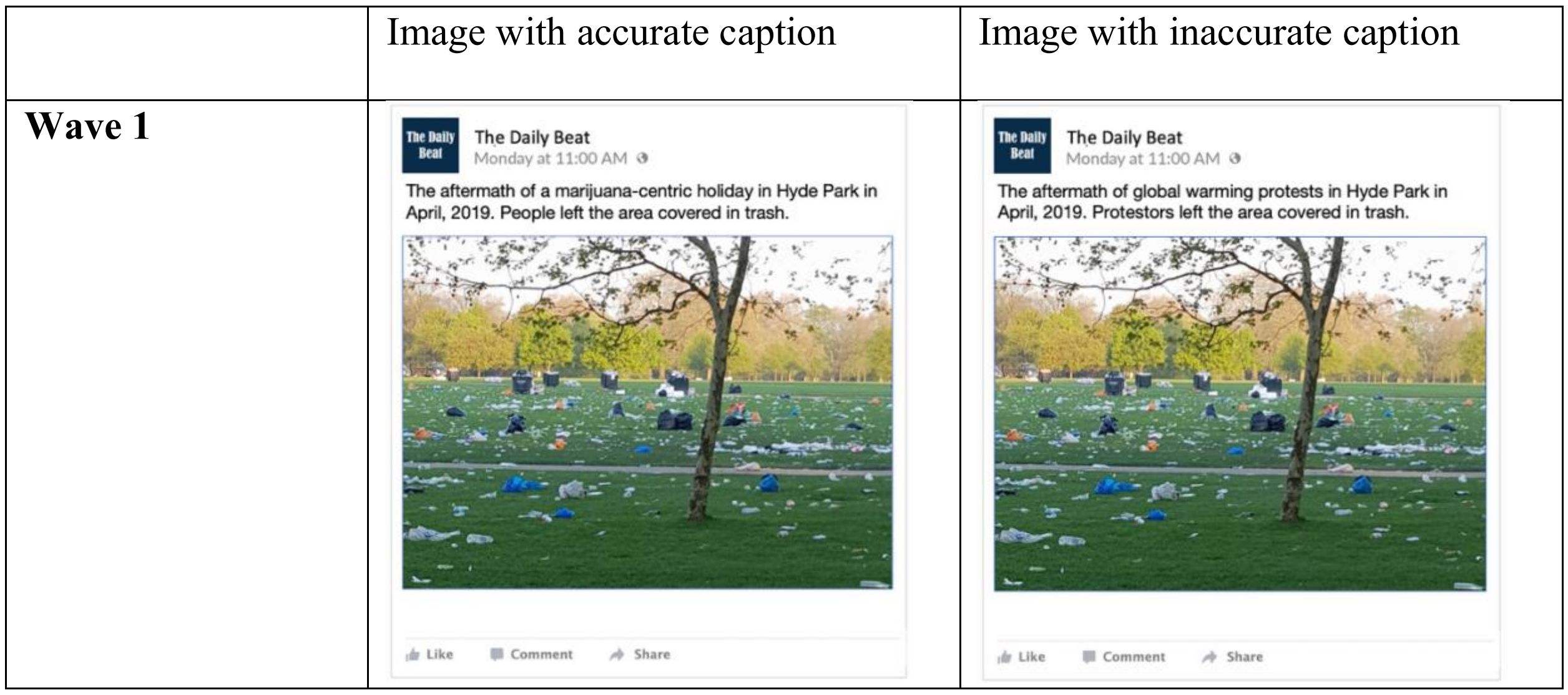

| | Image with accurate caption | Image with inaccurate caption |
|---|---|---|
| **Wave 1** | The Daily Beat, Monday at 11:00 AM: The aftermath of a marijuana-centric holiday in Hyde Park in April, 2019. People left the area covered in trash. | The Daily Beat, Monday at 11:00 AM: The aftermath of global warming protests in Hyde Park in April, 2019. Protestors left the area covered in trash. |

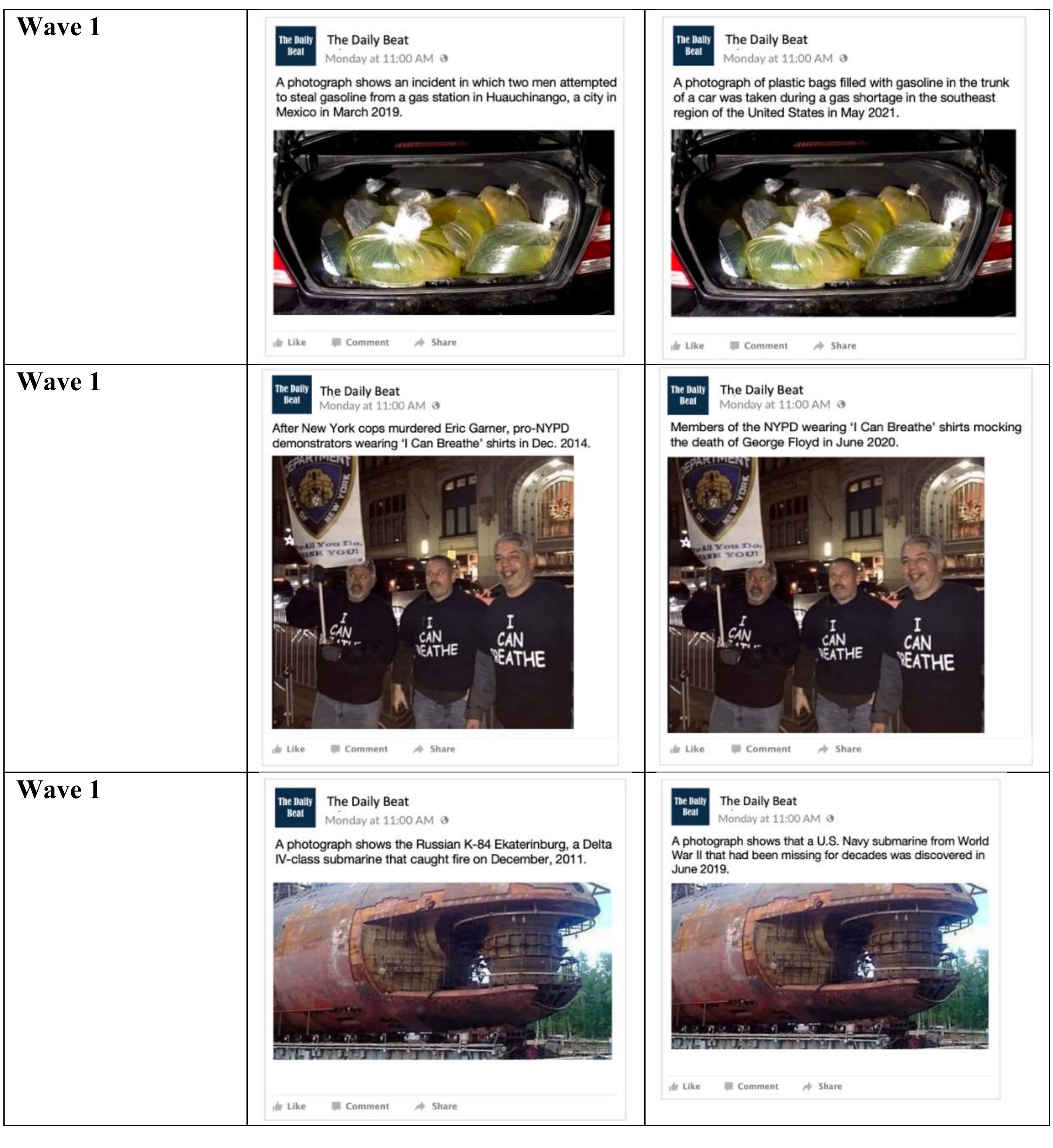

Wave 1
The Daily Beat
The Daily Beat
Monday at 11:00 AM
A photograph shows an incident in which two men attempted to steal gasoline from a gas station in Huauchinango, a city in Mexico in March 2019.
Like Comment Share
The Daily Beat
The Daily Beat
Monday at 11:00 AM
A photograph of plastic bags filled with gasoline in the trunk of a car was taken during a gas shortage in the southeast region of the United States in May 2021.
Like Comment Share
Wave 1
The Daily Beat
The Daily Beat
Monday at 11:00 AM
After New York cops murdered Eric Garner, pro-NYPD demonstrators wearing 'I Can Breathe' shirts in Dec. 2014.
I CAN BREATHE
Like Comment Share
The Daily Beat
The Daily Beat
Monday at 11:00 AM
Members of the NYPD wearing 'I Can Breathe' shirts mocking the death of George Floyd in June 2020.
I CAN BREATHE
Like Comment Share
Wave 1
The Daily Beat
The Daily Beat
Monday at 11:00 AM
A photograph shows the Russian K-84 Ekaterinburg, a Delta IV-class submarine that caught fire on December, 2011.
Like Comment Share
The Daily Beat
The Daily Beat
Monday at 11:00 AM
A photograph shows that a U.S. Navy submarine from World War II that had been missing for decades was discovered in June 2019.
Like Comment Share

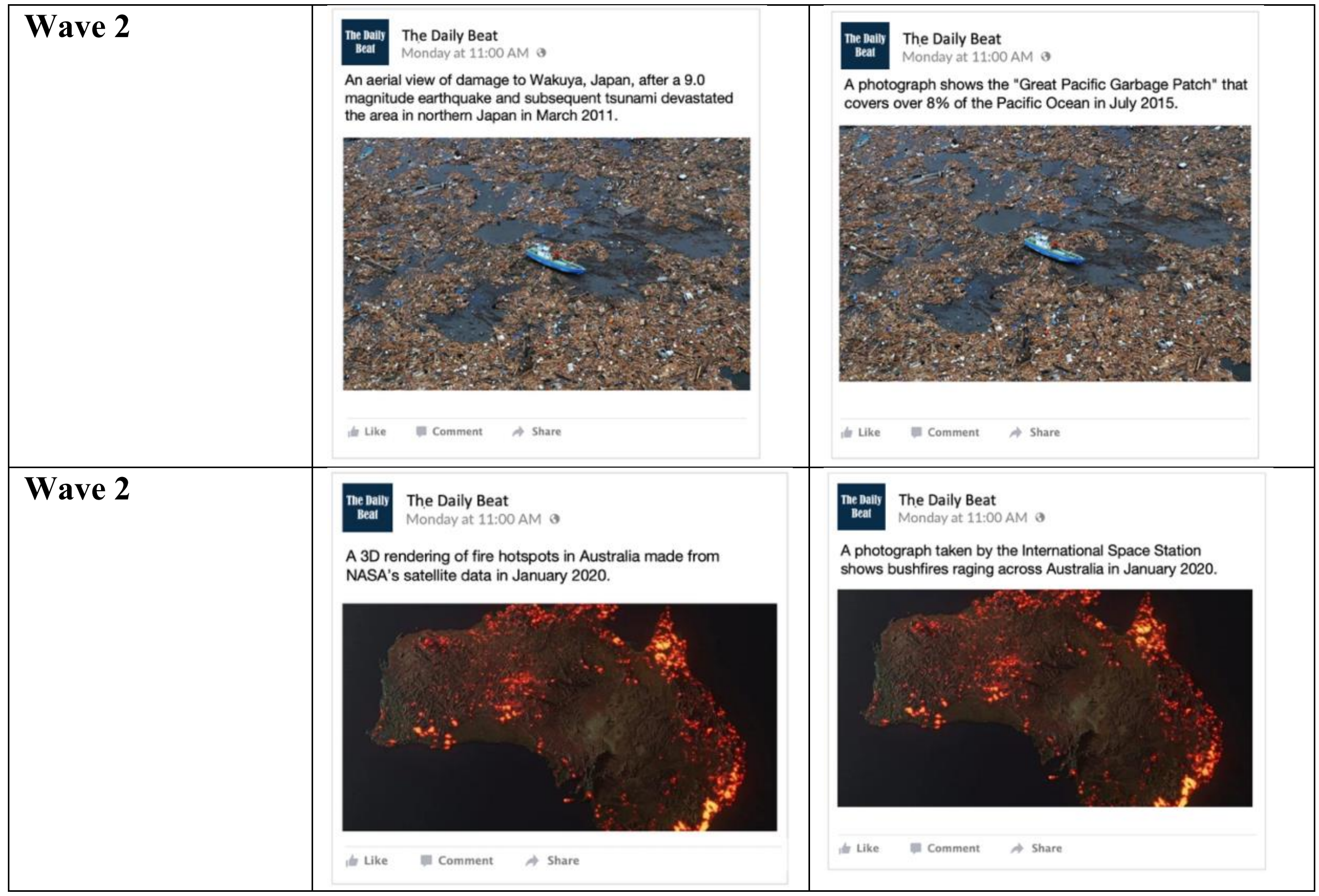
Wave 2
The Daily Beat
The Daily Beat
Monday at 11:00 AM
An aerial view of damage to Wakuya, Japan, after a 9.0 magnitude earthquake and subsequent tsunami devastated the area in northern Japan in March 2011.
Like
Comment
Share
The Daily Beat
The Daily Beat
Monday at 11:00 AM
A photograph shows the "Great Pacific Garbage Patch" that covers over 8% of the Pacific Ocean in July 2015.
Like
Comment
Share
Wave 2
The Daily Beat
The Daily Beat
Monday at 11:00 AM
A 3D rendering of fire hotspots in Australia made from NASA's satellite data in January 2020.
Like
Comment
Share
The Daily Beat
The Daily Beat
Monday at 11:00 AM
A photograph taken by the International Space Station shows bushfires raging across Australia in January 2020.
Like
Comment
Share

## Section 3: Measures

### *Main Outcome Variable*

**Discernment** was operationalized as participants' ability to differentiate between accurate and out-of-context news posts. For each participant, we first calculated the mean perceived credibility of accurate posts and the mean perceived credibility of out-of-context posts, then computed discernment as the difference between the two (credibility of accurate – credibility of out-of-context). Perceived accuracy of the news post was measured using four items. Participants were asked to indicate how credible they perceived each news post to be on a 5-point Likert scale. The items assessed whether the post was *credible*, *accurate*, and *reflected reality*, as well as whether the post *contained falsehoods* (reverse-coded). Responses were averaged to create a composite perceived accuracy score, with higher values indicating greater perceived accuracy. In Wave 1, participants viewed four news posts (two accurate, two out-of-context), and discernment was computed based on these four posts. In Wave 2, participants again rated the same four posts from Wave 1 and two additional ones (one accurate, one out-of-context). To capture participants' overall discernment at Wave 2, we used all available posts (six total) to compute the same difference score, averaging credibility judgments across all accurate and all out-of-context items respectively.

**Verification intention** was measured by assessing participants' intention to use reverse image search tools in the future. Participants were asked how likely they would be to use reverse image search when browsing news posts on social media, using a 5-point Likert scale ranging from *extremely unlikely* to *extremely likely*. Higher scores indicate a stronger intention to engage in image verification behaviors.

### *Covariates*

**Political ideology** was assessed using two items capturing participants' positions on social and economic issues, each rated on a 7-point scale from very liberal to very conservative (Akin et al., 2019). These items were highly correlated ($r = 0.85$) and averaged into a single index.

**Digital media literacy.** Participants reported their self-assessed understanding of nine internet-related concepts (e.g., advanced search, tagging, spyware, malware, phishing) on a 5-point scale ranging from 1 (No Understanding) to 5 (Full Understanding). This measure, adapted from Hargittai (2005), demonstrated excellent internal consistency (Cronbach's $\alpha = 0.93$). Responses were averaged to create a composite digital media literacy score.

**Visual literacy** was measured using four items evaluating participants' experience with digital and visual content creation tools (e.g., photography, image editing, digital illustration, and reverse image search). Responses were given on a 5-point scale (1 = No Experience; 5 = Expert/Professional). These items formed a reliable scale (Cronbach's $\alpha = 0.81$) and were averaged into a composite visual literacy score.

**Media trust** was measured by participants' agreement with two separate statements regarding their trust in mass media and social media to report news accurately, fairly, and completely (1 = Strongly Disagree; 5 = Strongly Agree; Guess et al., 2020). While the items were positively correlated ($r = 0.35$, $p < .001$), they represent distinct dimensions of trust and were therefore analyzed separately.

**Social media use** was captured by asking participants how frequently they used seven major platforms (Facebook, Twitter, Instagram, Snapchat, YouTube, Reddit, TikTok) on a 5-point scale (1 = Never; 5 = Always), with an option to add another platform. To quantify overall engagement, we counted the number of platforms for which participants reported moderate or higher usage (score $\geq 3$).

**Demographics.** Age (continuous), gender (female, male, and non-binary), education, and household income were included as control variables in the regression analyses. The median education level was a "College graduate (B.S., B.A., or other 4-year degree)", and the median income category was $60,000-69,999.

**Table S3.** Means, standard deviations, and correlations with confidence intervals among outcome variables and covariates

| Variable | *M* | *SD* | 1 | 2 | 3 | 4 | 5 | 6 | 7 | 8 | 9 | 10 |
|---|---|---|---|---|---|---|---|---|---|---|---|---|
| 1. Age | 42.20 | 14.12 | | | | | | | | | | |
| 2. Income | 6.64 | 3.19 | .04 | | | | | | | | | |
| | | | [-.02, .09] | | | | | | | | | |
| 3. Education | 5.38 | 1.24 | .12** | .38** | | | | | | | | |
| | | | [.07, .17] | [.33, .42] | | | | | | | | |
| 4. Media use | 4.54 | 1.78 | -.22** | .14** | .05 | | | | | | | |
| | | | [-.27, -.17] | [.09, .19] | [-.01, .10] | | | | | | | |
| 5. Political leaning | 3.24 | 1.82 | .19** | .08** | -.09** | -.03 | | | | | | |
| | | | [.14, .24] | [.03, .13] | [-.14, -.04] | [-.08, .03] | | | | | | |
| 6. Digital media literacy | 3.99 | 0.83 | .02 | .05 | .03 | .14** | -.06* | | | | | |

| Variable | M | SD | 1 | 2 | 3 | 4 | 5 | 6 | 7 | 8 | 9 | 10 |
|---|---|---|---|---|---|---|---|---|---|---|---|---|
| | | | [-.03, .07] | [-.00, .10] | [-.02, .08] | [.09, .20] | [-.11, -.00] | | | | | |
| 7. Visual literacy | 2.37 | 0.85 | -.18** | .05* | .02 | .31** | -.05* | .34** | | | | |
| | | | [-.23, -.13] | [.00, .11] | [-.03, .07] | [.26, .35] | [-.11, -.00] | [.29, .39] | | | | |
| 8. Discernment_W1 | 0.02 | 1.40 | .02 | -.00 | -.02 | -.01 | -.01 | -.00 | .03 | | | |
| | | | [-.03, .07] | [-.06, .05] | [-.07, .03] | [-.07, .04] | [-.07, .04] | [-.05, .05] | [-.02, .08] | | | |
| 9. Discernment_W2 | 0.68 | 1.15 | -.03 | .03 | .03 | -.02 | -.05 | .13** | .01 | -.05 | | |
| | | | [-.08, .03] | [-.03, .08] | [-.03, .09] | [-.07, .04] | [-.11, .01] | [.07, .19] | [-.05, .07] | [-.10, .01] | | |
| 10. Verification Intention_W1 | 3.90 | 1.16 | .03 | .02 | .02 | .16** | -.07* | .15** | .14** | .01 | .10** | |
| | | | [-.02, .09] | [-.03, .07] | [-.03, .07] | [.11, .21] | [-.12, -.01] | [.10, .20] | [.09, .19] | [-.04, .06] | [.05, .16] | |
| 11. Verific | 3.74 | 1.17 | -.00 | -.01 | -.01 | .18** | -.06* | .19** | .17** | .00 | .16** | .61** |
| ation Intention_W2 | | | | | | | | | | | | |
| | | | [-.06, .06] | [-.07, .05] | [-.07, .05] | [.12, .24] | [-.12, -.00] | [.13, .24] | [.11, .22] | [-.06, .06] | [.10, .21] | [.57, .65] |

*Note*. *M* and *SD* are used to represent mean and standard deviation, respectively. Values in square brackets indicate the 95% confidence interval for each correlation. The confidence interval is a plausible range of population correlations that could have caused the sample correlation (Cumming, 2014). * indicates $p < .05$. ** indicates $p < .01$.

**Section 4: Incentive Instruction for Each Experimental Group**

***Instructions: general (for all groups)***

UC Davis has developed a new social media site, and you're invited to be its first user.

***Instructions (Intervention * Result * Monetary):***

Note: In the following screens, you will see a total of 4 news posts shared on our social media site. The "Proceed" button will only appear after 10 seconds. For each post, you will be asked to evaluate the accuracy of the post. If your answers are correct, you will receive a BONUS PAYMENT of $1. In total, you can receive up to $4 bonus payment. Correct answers are based on fact-checkers. Your bonus payment will be delivered to your Prolific ID. Your answers to all other questions will not contribute to your bonus payment.

If you ever need a refresher on reverse image search, simply click the link located in the upper right corner.

***Instructions (Intervention * Task * Monetary):***

Note:  In the following screens, you will see a total of 4 news posts shared on our social media site. The "Proceed" button will only appear after 10 seconds. If you use reverse image search tools to evaluate the accuracy of each post, you will receive a BONUS PAYMENT of $1. In total, you can receive up to $4 bonus payment. Your bonus payment will be delivered to your Prolific ID.

If you ever need a refresher on reverse image search, simply click the link located in the upper right corner.

***Instructions (Intervention * Result *Symbolic):***

Note: In the following screens, you will see a total of 4 news posts shared on our social media site. The "Proceed" button will only appear after 10 seconds. For each post, you will be asked to evaluate the accuracy of the post. If your answers are correct, you will receive a verification badge. In total, you can receive up to 4 badges to be displayed in your profile. Correct answers are based on fact-checkers. Your answers to all other questions will not contribute to earning the badge.

If you ever need a refresher on reverse image search, simply click the link located in the upper right corner.

***Instructions (Intervention * Task * Symbolic):***

Note: In the following screens, you will see a total of 4 news posts shared on our social media site. The "Proceed" button will only appear after 10 seconds. If you use reverse image search

tools to evaluate the accuracy of each post, you will receive a verification badge. In total, you can receive up to 4 badges to be displayed in your profile. Your answers to all other questions will not contribute to earning the badge.

If you ever need a refresher on reverse image search, simply click the link located in the upper right corner.

***Instructions (Intervention * No Incentive):***

Note: In the following screens, you will see a total of 4 news posts shared on our social media site. The "Proceed" button will only appear after 10 seconds.

If you ever need a refresher on reverse image search, simply click the link located in the upper right corner.

***Instructions (No Intervention * No Incentive control):***

Note: In the following screens, you will see a total of 4 news posts shared on our social media site. The "Proceed" button will only appear after 10 seconds.

## Part II: Results Supplement

### Section 5: Group Descriptives

**Table S4.** Descriptive Means and Standard Errors for Discernment and Verification Intentions Across Incentive Conditions (Wave 1 and Wave 2)

| | Wave 1 | | | | | Wave 2 | | | | |
|---|---|---|---|---|---|---|---|---|---|---|
| | Discernment | | Verification Intention | | | Discernment | | Verification Intention | | |
| Group | Mean | SE | Mean | SE | N | Mean | SE | Mean | SE | N |
| Task * Symbolic | 0.07 | 0.082 | 4.09 | 0.067 | 236 | 0.54 | 0.062 | 3.89 | 0.067 | 189 |
| Task * Monetary | 0.25 | 0.097 | 4.34 | 0.056 | 235 | 0.75 | 0.083 | 4.12 | 0.064 | 190 |
| Result * Symbolic | -0.21 | 0.109 | 3.93 | 0.080 | 218 | 0.94 | 0.082 | 3.85 | 0.071 | 181 |
| Result * Monetary | 0.02 | 0.110 | 4.11 | 0.066 | 227 | 0.93 | 0.083 | 3.83 | 0.071 | 183 |
| No Incentive Control | -0.04 | 0.092 | 3.94 | 0.074 | 228 | 0.68 | 0.078 | 3.69 | 0.084 | 181 |
| | | | | | | | | | | |
| Task | 0.16 | 0.064 | 4.22 | 0.044 | 471 | 0.65 | 0.052 | 4.00 | 0.046 | 379 |
| Result | -0.10 | 0.078 | 4.02 | 0.052 | 445 | 0.94 | 0.058 | 3.84 | 0.050 | 364 |
| | | | | | | | | | | |
| Symbolic | -0.07 | 0.068 | 4.01 | 0.052 | 454 | 0.74 | 0.052 | 3.87 | 0.048 | 370 |
| Monetary | 0.14 | 0.073 | 4.23 | 0.043 | 462 | 0.84 | 0.059 | 3.98 | 0.048 | 373 |

## Section 6: Regression Analyses Results

To examine both the immediate and sustained effects of incentive mechanism and type, we estimated three regression models. Model 1 predicted Wave 1 outcomes, representing participants' discernment and intention immediately following the intervention when incentives were active. Model 2 predicted Wave 2 outcomes, one week later, after incentives were removed, capturing the persistence of treatment effects over time. Model 3 included Wave 1 outcomes as an additional covariate to account for individual baseline differences. This specification assesses whether the observed Wave 2 effects reflect genuine maintenance of improvement rather than pre-existing group differences.

**Table S5.** Regression Results Predicting Discernment by Incentive Mechanism and Type (Wave 1 and Wave 2)

| | *Dependent variable:* | | |
|---|---|---|---|
| | Discernment Wave 1 | Discernment Wave 2 | |
| | (1) | (2) | (3) |
| Age | 0.003 | $-0.008^{**}$ | $-0.008^{**}$ |
| | (0.003) | (0.003) | (0.003) |
| Gender Male | -0.017 | 0.105 | 0.103 |
| | (0.092) | (0.080) | (0.080) |
| Gender Nonbinary | -0.131 | 0.460 | 0.456 |
| | (0.339) | (0.277) | (0.277) |
| Income | 0.008 | 0.002 | 0.002 |
| | (0.015) | (0.013) | (0.013) |
| Education | -0.030 | 0.018 | 0.018 |
| | (0.040) | (0.034) | (0.034) |
| Media use | -0.012 | -0.020 | -0.021 |
| | (0.028) | (0.024) | (0.024) |
| Political leaning | -0.009 | -0.005 | -0.005 |
| | (0.027) | (0.023) | (0.023) |
| Digital media literacy | 0.005 | $0.201^{***}$ | $0.201^{***}$ |
| | (0.059) | (0.050) | (0.050) |
| Visual literacy | 0.107 | -0.057 | -0.055 |
| | (0.060) | (0.051) | (0.051) |
| Trust in mass media | 0.020 | $0.122^{***}$ | $0.123^{***}$ |
| | (0.041) | (0.035) | (0.035) |
| Trust in social media | -0.038 | -0.141*** | -0.142*** |
| | (0.045) | (0.038) | (0.038) |
| Task*Symbolic | 0.145 | -0.154 | -0.148 |
| | (0.141) | (0.121) | (0.121) |
| Task*Monetary | 0.298* | 0.112 | 0.118 |
| | (0.141) | (0.121) | (0.122) |
| Result*Symbolic | -0.149 | 0.271* | 0.268* |
| | (0.143) | (0.123) | (0.123) |
| Result*Monetary | 0.055 | 0.240 | 0.241* |
| | (0.142) | (0.123) | (0.123) |
| Discernment Wave 1 | | | -0.028 |
| | | | (0.025) |
| Constant | -0.240 | 0.268 | 0.256 |
| | (0.371) | (0.320) | (0.320) |
| Observations | 1,116 | 907 | 907 |
| $R^2$ | 0.016 | 0.071 | 0.072 |
| Adjusted $R^2$ | 0.003 | 0.055 | 0.056 |
| Residual Std. Error | 1.492 (df = 1100) | 1.152 (df = 891) | 1.152 (df = 890) |
| F Statistic | 1.200 (df = 15; 1100) | 4.547*** (df = 15; 891) | 4.340*** (df = 16; 890) |
| *Note:* | | | *p<0.05; **p<0.01; ***p<0.001 |

**Table S6.** Regression Results Predicting Verification Intention by Incentive Mechanism and Type (Wave 1 and Wave 2)

| | *Dependent variable:* | | |
|---|---|---|---|
| | Verification Intention Wave 1 | Verification Intention Wave 2 | |
| | (1) | (2) | (3) |
| Age | $0.006^{**}$ | 0.001 | -0.003 |
| | (0.002) | (0.003) | (0.002) |
| Gender Male | -0.065 | $-0.164^{*}$ | $-0.139^{*}$ |
| | (0.062) | (0.071) | (0.061) |
| Gender Nonbinary | 0.215 | 0.338 | 0.197 |
| | (0.228) | (0.247) | (0.211) |
| Income | 0.003 | -0.011 | -0.011 |
| | (0.010) | (0.012) | (0.010) |
| Education | -0.021 | -0.050 | -0.044 |
| | (0.027) | (0.030) | (0.026) |
| Media use | $0.071^{***}$ | $0.076^{***}$ | 0.026 |
| | (0.019) | (0.021) | (0.018) |
| Political leaning | -0.019 | 0.001 | 0.011 |
| | (0.018) | (0.021) | (0.018) |
| Digital media literacy | $0.212^{***}$ | $0.252^{***}$ | $0.136^{***}$ |
| | (0.039) | (0.045) | (0.039) |
| Visual literacy | 0.041 | $0.120^{**}$ | $0.118^{**}$ |
| | (0.040) | (0.046) | (0.039) |
| Trust in mass media | $0.085^{**}$ | $0.097^{**}$ | 0.042 |
| | (0.027) | (0.032) | (0.027) |
| Trust in social media | $-0.073^{*}$ | $-0.104^{**}$ | $-0.066^{*}$ |
| | (0.030) | (0.034) | (0.029) |
| Task*Symbolic | 0.144 | 0.209 | 0.151 |
| | (0.095) | (0.108) | (0.092) |
| Task*Monetary | $0.411^{***}$ | $0.429^{***}$ | $0.207^{*}$ |
| | (0.095) | (0.108) | (0.093) |
| Result*Symbolic | -0.010 | 0.165 | $0.201^{*}$ |
| | (0.096) | (0.109) | (0.093) |
| Result*Monetary | 0.178 | 0.126 | 0.048 |
| | (0.096) | (0.109) | (0.093) |
| Intention Wave 1 | | | $0.529^{***}$ |
| | | | (0.029) |
| Constant | $2.520^{***}$ | $2.413^{***}$ | $1.197^{***}$ |
| | (0.249) | (0.285) | (0.252) |
| Observations | 1,116 | 907 | 907 |
| $R^2$ | 0.099 | 0.122 | 0.361 |
| Adjusted $R^2$ | 0.087 | 0.107 | 0.350 |
| Residual Std. Error | 1.002 (df = 1100) | 1.026 (df = 891) | 0.876 (df = 890) |
| F Statistic | $8.086^{***}$ (df = 15; 1100) | $8.265^{***}$ (df = 15; 891) | $31.492^{***}$ (df = 16; 890) |
| *Note:* | | | $^{*}p<0.05$; $^{**}p<0.01$; $^{***}p<0.001$ |

## Section 7: Additional Pre-registered Analyses

In addition to the primary analyses, the pre-registration specified several secondary outcomes related to participants' perceptions and behaviors, including perceived credibility of accurately attributed visual posts, perceived credibility of misattributed visual posts, confidence in credibility judgments, perceived self-efficacy in identifying misinformation, and engagement in reverse image search behaviors. While these outcomes were not the primary focus of the main analyses, they were included in the pre-registration to provide a more comprehensive assessment of the intervention's effects. This section reports the results of these pre-registered secondary analyses.

### Measures

*Perceived accuracy* of each news post was measured using a four-item scale assessing participants' evaluations of the credibility of the visual post. After viewing each post, participants indicated their agreement with the following statements on a 5-point scale: *The post was credible*; *The post was accurate*; *The post reflected reality*; and *The post contained falsehoods* (reverse coded). Responses were averaged to create a composite perceived accuracy score, with higher values indicating greater perceived accuracy.

*Confidence in credibility judgment* was measured immediately after participants completed the perceived accuracy ratings for each post. Participants were asked: *"How confident are you in your answer to the previous question about the credibility of the post?"* Responses were recorded on a 5-point scale ranging from *not at all confident* to *extremely confident*.

*Perceived self-efficacy in identifying misinformation* was measured using a five-item scale adapted from prior work (e.g., Chung & Wihbey, 2022). Participants indicated the extent to which they agreed with the following statements on a 5-point scale: *I can identify misinformation if I read it*; *I can tell when a news story contains misinformation*; *It is easy for me to determine what information is trustworthy*; *I can identify the source of images*; and *I know how to use reverse image search tools*. Items were averaged to form a composite self-efficacy score, with higher values indicating greater perceived efficacy.

In addition to self-reported verification intentions, we included an observational behavioral measure capturing whether participants right-clicked on an image while viewing each post[1]. Right-clicking an image enables access to reverse image search functionality in common web browsers and therefore serves as a behavioral proxy for engagement in image verification. We note that this measure is necessarily an approximation: while a right-click indicates an opportunity to initiate reverse image search, we cannot directly observe whether participants subsequently selected a reverse image search option or used the action for other purposes (e.g., saving the image). Nonetheless, relative to self-reported measures, right-click behavior provides a more conservative and behaviorally grounded indicator of verification-related engagement.

[1] Due to a technical issue in the professionally designed social media platform, right-click interaction data were not recorded for participants assigned to the control condition in Wave 1. As a result, analyses of observed right-click behavior are limited to participants in the intervention conditions, and direct comparisons between intervention and control groups on this behavioral outcome are not possible. We report these results for completeness and interpret them with appropriate caution.

**Table S7.** Regression Results Predicting Perceived Credibility of Accurately Attributed Visual Posts by Incentive Mechanism and Type (Wave 1 and Wave 2)

| | *Dependent variable:* | | |
|---|---|---|---|
| | Perceived Accuracy of Accurately Attributed Wave 1 | Perceived Accuracy of Accurately Attributed Wave 2 | |
| | (1) | (2) | (3) |
| Age | 0.001 | -0.007** | -0.007** |
| | (0.002) | (0.002) | (0.002) |
| Gender Male | 0.026 | 0.022 | 0.024 |
| | (0.064) | (0.062) | (0.060) |
| Gender Nonbinary | 0.061 | 0.446* | 0.429* |
| | (0.235) | (0.214) | (0.207) |
| Income | 0.005 | -0.003 | -0.003 |
| | (0.011) | (0.010) | (0.010) |
| Education | -0.013 | -0.003 | -0.002 |
| | (0.028) | (0.026) | (0.025) |
| Media use | 0.0001 | -0.039* | -0.038* |
| | (0.019) | (0.019) | (0.018) |
| Political leaning | 0.013 | 0.006 | 0.004 |
| | (0.019) | (0.018) | (0.017) |
| Media literacy | -0.034 | 0.054 | 0.059 |
| | (0.041) | (0.039) | (0.037) |
| Visual literacy | 0.160*** | 0.031 | -0.0005 |
| | (0.041) | (0.040) | (0.038) |
| Trust in mass media | 0.015 | 0.070* | 0.061* |
| | (0.028) | (0.027) | (0.026) |
| Trust in social media | 0.098** | -0.012 | -0.032 |
| | (0.031) | (0.030) | (0.029) |
| Task*Symbolic | 0.035 | -0.067 | -0.088 |
| | (0.098) | (0.094) | (0.091) |
| Task*Monetary | 0.132 | -0.0002 | -0.029 |
| | (0.098) | (0.094) | (0.091) |
| Result*Symbolic | -0.096 | 0.171 | 0.189* |
| | (0.099) | (0.095) | (0.092) |
| Result*Monetary | 0.041 | 0.212* | 0.205* |
| | (0.099) | (0.095) | (0.092) |
| Perceived Accuracy Wave 1 | | | 0.220*** |
| | | | (0.027) |
| Constant | 2.264*** | 3.170*** | 2.679*** |
| | (0.257) | (0.247) | (0.247) |
| Observations | 1,116 | 907 | 907 |
| $R^2$ | 0.038 | 0.044 | 0.108 |
| Adjusted $R^2$ | 0.025 | 0.028 | 0.092 |
| Residual Std. Error | 1.035 (df = 1100) | 0.890 (df = 891) | 0.860 (df = 890) |
| F Statistic | 2.916*** (df = 15; 1100) | 2.740*** (df = 15; 891) | 6.769*** (df = 16; 890) |
| *Note:* | | | *p<0.05; **p<0.01; ***p<0.001 |

**Table S8.** Regression Results Predicting Perceived Credibility of Misattributed Visual Posts, perceived by Incentive Mechanism and Type (Wave 1 and Wave 2)

| | *Dependent variable:* | | |
|---|---|---|---|
| | Perceived Accuracy of Misattributed Wave 1 | Perceived Accuracy of Misattributed Wave 2 | |
| | (1) | (2) | (3) |
| Age | -0.003 | 0.001 | 0.002 |
| | (0.002) | (0.002) | (0.002) |
| Gender Male | 0.044 | -0.083 | -0.091 |
| | (0.066) | (0.059) | (0.057) |
| Gender Nonbinary | 0.192 | -0.014 | -0.058 |
| | (0.243) | (0.204) | (0.199) |
| Income | -0.003 | -0.004 | -0.004 |
| | (0.011) | (0.010) | (0.009) |
| Education | 0.017 | -0.021 | -0.024 |
| | (0.028) | (0.025) | (0.024) |
| Media use | 0.012 | -0.019 | -0.019 |
| | (0.020) | (0.018) | (0.017) |
| Political leaning | 0.022 | 0.011 | 0.009 |
| | (0.020) | (0.017) | (0.017) |
| Media literacy | -0.039 | $-0.146^{***}$ | $-0.142^{***}$ |
| | (0.042) | (0.037) | (0.036) |
| Visual literacy | 0.052 | $0.088^{*}$ | $0.078^{*}$ |
| | (0.043) | (0.038) | (0.037) |
| Trust in mass media | -0.006 | $-0.052^{*}$ | $-0.052^{*}$ |
| | (0.029) | (0.026) | (0.025) |
| Trust in social media | $0.136^{***}$ | $0.129^{***}$ | $0.109^{***}$ |
| | (0.032) | (0.028) | (0.028) |
| Task*Symbolic | -0.110 | 0.087 | 0.103 |
| | (0.101) | (0.089) | (0.087) |
| Task*Monetary | -0.165 | -0.113 | -0.098 |
| | (0.101) | (0.090) | (0.087) |
| Result*Symbolic | 0.053 | -0.100 | -0.106 |
| | (0.102) | (0.091) | (0.088) |
| Result*Monetary | -0.014 | -0.028 | -0.031 |
| | (0.102) | (0.090) | (0.088) |
| Perceived Accuracy Wave 1 | | | $0.181^{***}$ |
| | | | (0.026) |
| Constant | $2.504^{***}$ | $2.902^{***}$ | $2.422^{***}$ |
| | (0.266) | (0.236) | (0.239) |
| Observations | 1,116 | 907 | 907 |
| $R^2$ | 0.037 | 0.062 | 0.112 |
| Adjusted $R^2$ | 0.024 | 0.046 | 0.096 |
| Residual Std. Error | 1.069 (df = 1100) | 0.849 (df = 891) | 0.826 (df = 890) |
| F Statistic | $2.799^{***}$ (df = 15; 1100) | $3.895^{***}$ (df = 15; 891) | $7.008^{***}$ (df = 16; 890) |
| *Note:* | | | $^{*}p<0.05$; $^{**}p<0.01$; $^{***}p<0.001$ |

**Table S9.** Regression Results Predicting Confidence in Credibility Judgment by Incentive Mechanism and Type (Wave 1 and Wave 2)

| | *Dependent variable:* | | |
|---|---|---|---|
| | Confidence Wave 1 | Confidence Wave 2 | |
| | (1) | (2) | (3) |
| Age | 0.001 | 0.002 | 0.002 |
| | (0.002) | (0.002) | (0.002) |
| Gender Male | $0.113^{*}$ | $0.168^{**}$ | $0.133^{**}$ |
| | (0.056) | (0.056) | (0.046) |
| Gender Nonbinary | 0.161 | 0.213 | 0.132 |
| | (0.204) | (0.194) | (0.161) |
| Income | 0.001 | 0.002 | 0.001 |
| | (0.009) | (0.009) | (0.008) |
| Education | $-0.050^{*}$ | -0.009 | 0.015 |
| | (0.024) | (0.024) | (0.020) |
| Media use | $0.038^{*}$ | $0.038^{*}$ | 0.010 |
| | (0.017) | (0.017) | (0.014) |
| Political leaning | -0.017 | 0.003 | 0.013 |
| | (0.016) | (0.016) | (0.013) |
| Media literacy | $0.271^{***}$ | $0.223^{***}$ | $0.091^{**}$ |
| | (0.035) | (0.035) | (0.030) |
| Visual literacy | 0.005 | 0.059 | $0.060^{*}$ |
| | (0.036) | (0.036) | (0.030) |
| Trust in mass media | $0.077^{**}$ | $0.074^{**}$ | 0.029 |
| | (0.025) | (0.025) | (0.021) |
| Trust in social media | -0.003 | 0.008 | 0.016 |
| | (0.027) | (0.027) | (0.022) |
| Task*Symbolic | $0.171^{*}$ | -0.119 | $-0.167^{*}$ |
| | (0.085) | (0.085) | (0.070) |
| Task*Monetary | $0.477^{***}$ | 0.156 | -0.073 |
| | (0.085) | (0.085) | (0.071) |
| Result*Symbolic | $0.380^{***}$ | -0.001 | $-0.178^{*}$ |
| | (0.086) | (0.086) | (0.072) |
| Result*Monetary | 0.419*** | 0.057 | -0.132 |
| | (0.086) | (0.086) | (0.072) |
| Confidence Wave 1 | | | 0.503*** |
| | | | (0.025) |
| Constant | 2.140*** | 1.695*** | 0.586** |
| | (0.224) | (0.224) | (0.193) |
| Observations | 1,116 | 904 | 904 |
| $R^2$ | 0.125 | 0.118 | 0.395 |
| Adjusted $R^2$ | 0.113 | 0.103 | 0.384 |
| Residual Std. Error | 0.901 (df = 1100) | 0.805 (df = 888) | 0.667 (df = 887) |
| F Statistic | 10.430*** (df = 15; 1100) | 7.943*** (df = 15; 888) | 36.163*** (df = 16; 887) |
| *Note:* | | | *p<0.05; **p<0.01; ***p<0.001 |

**Table S10.** Regression Results Predicting Self-efficacy in Identifying Misinformation by Incentive Mechanism and Type (Wave 1 and Wave 2)

| | *Dependent variable:* | | |
|---|---|---|---|
| | Self-Efficacy Wave 1 | Self-Efficacy Wave 2 | |
| | (1) | (2) | (3) |
| Age | -0.005*** | -0.006** | -0.002 |
| | (0.002) | (0.002) | (0.001) |
| Gender Male | 0.086* | 0.108* | 0.065 |
| | (0.042) | (0.047) | (0.036) |
| Gender Nonbinary | 0.019 | 0.203 | 0.190 |
| | (0.154) | (0.162) | (0.124) |
| Income | 0.010 | 0.010 | 0.004 |
| | (0.007) | (0.008) | (0.006) |
| Education | -0.012 | -0.017 | -0.013 |
| | (0.018) | (0.020) | (0.015) |
| Media use | 0.026* | 0.037** | 0.016 |
| | (0.013) | (0.014) | (0.011) |
| Political leaning | -0.018 | -0.015 | -0.003 |
| | (0.012) | (0.013) | (0.010) |
| Media literacy | 0.217*** | 0.220*** | 0.078*** |
| | (0.027) | (0.029) | (0.023) |
| Visual literacy | 0.079** | 0.073* | 0.031 |
| | (0.027) | (0.030) | (0.023) |
| Trust in mass media | 0.010 | 0.028 | 0.018 |
| | (0.019) | (0.021) | (0.016) |
| Trust in social media | -0.004 | 0.001 | -0.001 |
| | (0.020) | (0.022) | (0.017) |
| Task*Symbolic | 0.201** | 0.040 | -0.077 |
| | (0.064) | (0.071) | (0.055) |
| Task*Monetary | 0.326*** | 0.177* | -0.016 |
| | (0.064) | (0.071) | (0.055) |
| Result*Symbolic | 0.131* | 0.069 | -0.023 |
| | (0.065) | (0.072) | (0.055) |
| Result*Monetary | 0.155* | 0.052 | -0.028 |
|  | (0.065) | (0.072) | (0.055) |
| Self-Efficacy Wave 1 |  |  | 0.630*** |
|  |  |  | (0.025) |
| Constant | 2.744*** | 2.768*** | 1.082*** |
|  | (0.168) | (0.187) | (0.159) |
| Observations | 1,116 | 907 | 907 |
| $R^2$ | 0.152 | 0.158 | 0.503 |
| Adjusted $R^2$ | 0.140 | 0.143 | 0.494 |
| Residual Std. Error | 0.676 (df = 1100) | 0.672 (df = 891) | 0.516 (df = 890) |
| F Statistic | 13.106*** (df = 15; 1100) | 11.114*** (df = 15; 891) | 56.309*** (df = 16; 890) |
| *Note:* |  |  | *p<0.05; **p<0.01; ***p<0.001 |

**Table S11.** Observed Right-Click Behavior (Descriptive).

| Condition | Wave 1 | | | | Wave 2 | | | |
|---|---|---|---|---|---|---|---|---|
| | **Mean RC (0–4)** | **% Zero** | **% All** | **N** | **Mean RC (0–6)** | **% Zero** | **% All** | **N** |
| No Incentive | - | - | - | - | 2.05 | 40.0% | 17.1% | 181 |
| Task × Symbolic | 3.36 | 8.1% | 72.0% | 236 | 2.01 | 41.3% | 16.4% | 189 |
| Task × Monetary | 3.65 | 6.0% | 86.0% | 235 | 2.31 | 34.7% | 22.6% | 190 |
| Result × Symbolic | 2.63 | 26.6% | 59.2% | 218 | 1.73 | 42.5% | 13.3% | 181 |
| Result × Monetary | 2.66 | 26.9% | 57.3% | 227 | 2.06 | 41.0% | 19.7% | 183 |

Note: Because right-click data were not recorded for control participants in Wave 1, these results are intended to document behavioral patterns within intervention conditions rather than estimate causal effects.